\documentclass[prd,twocolumn,groupedaddress,showpacs]{revtex4-1}
\usepackage[utf8]{inputenc}
\usepackage[T1]{fontenc}
\usepackage{graphics}	
\usepackage{amsmath}	
\usepackage[unicode=true,bookmarks=true,bookmarksnumbered=false,bookmarksopen=false, breaklin
ks=false,pdfborder={0 0 0},backref=false,colorlinks=true,citecolor=red,linkcolor
=blue,urlcolor=blue]{hyperref}
\newcommand{\orcid}[1]{\href{https://orcid.org/#1}{\resizebox{10px}{!}{\includegraphics{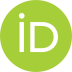}}}}

\begin{document}
\title{Probing spherical outbursts of neutron star mergers from an equation of
state with composite nucleons}

\author{Vikram Soni\orcid{0000-0001-9900-3495}}
\email{vsoni.physics@gmail.com}
\affiliation{Centre for Theoretical Physics, Jamia Millia Islamia, New Delhi - 110025.
}

\date{Accepted XXX. Received YYY; in original form ZZZ}


\begin{abstract}
We consider a variational 'crystalline' equation of state (EOS) that follows from
nucleons that are chiral solitons with deep relativistic quark bound states. In
this model conventional quark matter does not occur till a high density
threshold at which the quark bound states in the nucleons get compressed
and merge with the continuum. Once the barrier at this threshold is overcome,
roughly when, $n_B \sim 1/fm^3$, we expect the 'crystalline' nuclear matter to make a
sudden transition into quark matter when the EOS becomes soft through a
decompression. The sudden increase in density triggers a collapse releasing
large amount of gravitational potential energy that can generate a spherical
outburst (or kilonova) of ejected matter.
\end{abstract}

\maketitle



\section{Introduction}

The merger of two neutron stars which resulted in a clear signal of
gravitational wave emission, GW170817, was reported in
2017 \citep{GW170817}.
In this
letter we look at the surprising and unexpectedly almost perfectly spherical burst of ejecta
that has been deduced from the data analysis of this event by
Sneppen et al. \citep{Sneppen_2023}. According to these authors, this
is so as most of the previous analysis of the merger of two rapidly
rotating neutron stars  suggests an anisotropic, possibly disc like mass
distribution of the merging state and the ejecta.
Working with composite nucleons that are chiral solitons with deep
relativistic quark bound states \citep{Kahana_Ripka_Soni_1984,Birse_Banerjee_1984}  led us to a distinctive equation
of state (EOS) \citep{Banerjee_Glendenning_Soni_1985} for nuclear matter . This letter employs  earlier work
\citep{Soni_2019}, in  the context of this new finding \citep{Sneppen_2023}, that could potentially
provide an explanation of the property of the spherical  outburst.

At the outset we have to say that  exact calculations of the ground states
are very complicated and daunting as they involve not only solitons but
with quark bound states.   Though our considerations are approximate they
point to a new EOS for neutron stars that could provide an insight  in
understanding  this puzzle.

\section{Dense Solitonic `Crystalline' Nuclear Matter}

One significant difference with most of the equations of state in the literature
and this work is that we have composite nucleons, with deeply bound quarks,
where the structure of the nucleon plays an essential role in the transition
from nuclear matter to quark matter at high density.

We find that at densities above the density when nucleons overlap
the deeply bound quarks remain bound till much higher density producing a
hard core interaction between the nucleons. The deeper the bound states the
greater the hard core interaction (sec. IV of \citet{Soni_2019}).

One plausible but approximate way to incorporate this hard core aspect for
dense matter, at baryon densities beyond nucleon overlap, is through a
variational 'crystalline' (in a single cell Wigner Sietz approximation) ground
state .Such a calculation that was carried out by Banerjee, Glendenning and
Soni \citep{Banerjee_Glendenning_Soni_1985} with some interesting findings. This is a relativistic( Dirac) band
structure calculation of a cubic lattice of solitonic composite nucleons, with
quarks bound in a skyrme soliton background.

In this model conventional quark matter does not occur till a high density
threshold at which the quark bound states in the nucleons get compressed
and merge with the continuum. Another way to understand this 
phenomenon is to think of the deep bound state quarks as providing a hard
core potential barrier( in analogy to a coulomb barrier) between the solitonic
crystal phase and the far softer normal quark matter phase -the barrier is
overcome at the threshold density \citep{Banerjee_Glendenning_Soni_1985}.
We note that our calculations, though quite approximate, point to a new and distinct
equation of state.
More detail can be found in \citet{Soni_2019} and references therein.

\section{Transition from 'Crystalline' Nuclear Matter to Conventional Quark Matter}

In \citet{Soni_2019} we made a rudimentary estimate for the 'crystalline' state EOS of our
composite solitons and found that upto the threshold density it can be
potentially similar to the widely accepted APR 98 \citep{PhysRevC.58.1804}) nuclear EOS which carries a hard core
interaction between elementary ( not composite) nucleons. We would like to
emphasize that the composite solitonic crystal state is a new state of matter
that is neither conventional quark matter nor conventional nuclear matter.
However,  there is an important difference in the behaviour at the phase
transition to quark matter, between our model and the APR 98 EOS. The
APR phase can transit into  a mixed phase at constant pressure before
going entirely into the quark matter phase. The mixed phase lasts till
a nucleon density, $n_B  < 0.8-0.6/fm^3$ before exiting into a purely quark
matter ( neutral pion condensed) phase as can be seen from the Maxwell
construction (Fig1 of \citet{Soni_2019,Soni_Bhattacharya_2006}).
In our crystal phase the quarks
remain bound till higher density excluding such a mixed phase – thus
going directly, through a sudden transition, to quark matter.

As indicated in \citet{Banerjee_Glendenning_Soni_1985,Soni_2019} this happens roughly when $n_B \sim 1/fm^3$. Once the
barrier at this threshold is overcome, we expect our nuclear matter
to make a sharp  transition into  quark matter. This is the point when
the EOS becomes soft through a decompression. The sudden increase in
density can result in a collapse generating a burst
which ejects matter.

Thus we are led to the conclusion that in the alternative model we have presented,
all neutron stars should have no regular quark matter but have a nuclear core that can
be represented by the `crystalline' state above. Neutron stars with masses
over the maximum mass $M_{max}$, or for central nucleon density larger than
the threshold density,
$n_B \sim 1/fm^3$, will become unstable and transit to quark matter and then
collapse. For neutron
star mergers this would occur in the later stage when their cores overlap.

\section{Properties of the Outburst of Matter in the Post Transition Collapse}

The contraction of the core due to the sharp change in the compressibility of EOS
at the threshold density of our model would result in a different post merger
scenario, in contrast to hybrid crossover models \citep{Baym_Hatsuda_Kojo_Powell_Song_Takatsuka_2018} where there is no such
effect ( density discontinuity). This may be observable.

The sudden phase transition from the stiff relativistic crystalline (quark
soliton) state to the soft conventional quark matter will result in a
decompression as the pressure decreases markedly in the quark matter phase
at the threshold density.
\begin{figure}
\resizebox{8.0cm}{!}{\includegraphics{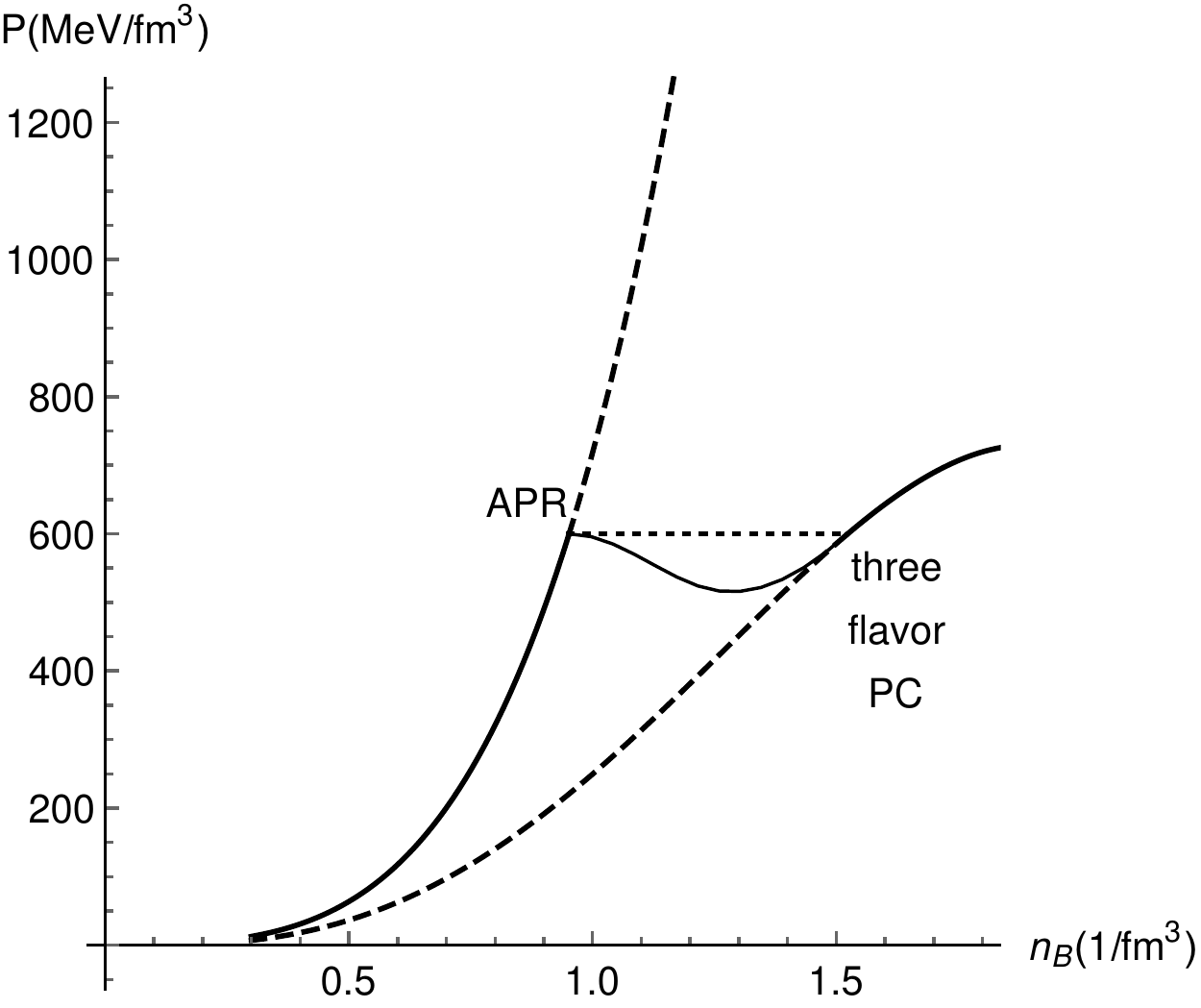}}
\caption{Pressure, $P$ ($Mev/fm^3$) vs $n_B$ ($1/fm^3$), for the APR EOS, and the
pion condensed 3 flavour quark matter. Illustration of the transition that
occurs at, $n_B (0.95/fm^3$), in the nuclear phase. The solid line indicates
the decompression at  the phase transition (From Fig 5 in \citet{Soni_2019}).}
\label{fig1}
\end{figure}

As we have indicated earlier, the APR 98 nuclear
matter EOS \citep{PhysRevC.58.1804} is similar to our EOS
\citep{Banerjee_Glendenning_Soni_1985}  upto the threshold density. The
APR  98 is a much used and comprehensive nuclear EOS compared to our
rather crude  and heuristic model. Thus, to simply
illustrate this decompression we calculate the pressure and gravitational
potential energy loss using the APR nuclear matter EOS, only as a
substitute for our model up to the threshold density. The transition from
the APR 98  EOS  to the quark matter (pion condensed) EOS is given in
Fig. \ref{fig1}. If the sudden phase transition to 
quark matter occurs at baryon density around, $n_B \sim 0.95/fm^3$, we find
that the pressure in the nuclear APR 98 phase at this density is, $P \sim
600~\text{Mev}/fm^3$.

Of course, the pressure, P, and the energy per baryon, EB, in the quark matter
state at this density are much lower. Since we need to balance the pressure in
both phases it is pertinent to find the density at which the same pressure
occurs in the quark matter’ state. From Figure \ref{fig1}, this is found to be, $n_B \sim
1.5/fm^3$, and the corresponding, $EB \sim 1260$ MeV. Thus, as nuclear matter
clears the threshold barrier set by the soliton crystal and goes into quark matter, there will be a sudden
contraction followed by a consequent increase in baryon density from,
${n_B}_1 \sim 0.95/fm^3$ to ${n_B}_2 \sim 1.5/fm^3$, when the system attains
the same pressure.

In a high mass ‘compound’ neutron star or in the merger of 2 neutron stars
such a major change in compressibility, $K$, would cause a contraction of the
core and bring down the gravitational potential energy of the star. A rough
estimate( Newtonian) of the gravitational energy release is provided by
considering a neutron star of mass $M$ whose gravitational potential energy is,
$\tfrac{3}{5}GM^2/R$. Keeping the mass fixed we can write down the energy
difference, $\Delta E_G$, as we change the mass density indicated above 
from $\rho_1 \sim 1.6 \times  10^{15}$ gm/cc ($n_{B1} \sim 0.95/fm^3$) to
$\rho_2 \sim 2.53 \times 10^{15}$ gm/cc ($n_{B2} \sim 1.5/fm^3$), for a uniform
density star, where, $R = (3M/4\pi\rho)^{1/3}$

\begin{equation}
\Delta E_G = \tfrac{3}{5}GM^2 (1/R_2 - 1/R_1).
\end{equation}

On substituting the values of a 2 solar mass star, $M \sim 4 \times 10^{33}$ gm
and the
above density change for the corresponding radii, we can get a sudden release
of gravitational energy of $\Delta E_G \sim 0.7\times 10^{53}$ ergs. This can yield a matter
ejecting burst. This is of the order of energy injection, tens of Mev/nucleon,
postulated by Sneppen et al \citep{Sneppen_2023} to get a spherical burst.

In \citet{Sneppen_2023} their simulation in the  Methods section on ‘Numerical models
of late-time energy injection’ , suggests that   most of the energy
is released within the first second of post-merger evolution. The models
reveal that an energy injection ‘of more than about 30 MeV per baryon
(see eqn. 6 in \citet{Sneppen_2023}) would be necessary to induce a nearly spherical density
distribution at velocities greater than $0.2 c$ (left panel of Extended
Data Figure 4)’.  They find that such high values exclude radioactive
heating  and several magnetic energy sources being  viable scenarios
to provide the high degree of sphericities  of  the ejecta .

At this point it is important  to recall  that in our model the outburst
would follow  in the final stages  when the cores of the stars come
together.  This is therefore a central explosion which is most likely
to  yield an  isotropic (spherical) outburst.
In  contrast  the rotational energy dissipation or the magnetic energy loss, from magnetar strength magnetic fields, fall short
of the required energy for the outburst  above. Besides, they 
are intrinsically directional and anisotropic.

\section{Conclusion}

Though our estimates are approximate, they present a new and plausible scenario
for neutron stars and their merger. What we have pointed out that in
our model once the hard core barrier provided by a crystalline nuclear
state is overcome at the threshold density,  it can rapidly transform
to quark matter, which has a much softer EOS. This could well result in
a decompression accompanied by a high energy ‘spherical’ burst as
the cores of the two  neutron stars merge.

\section*{Acknowledgements}
We are happy to acknowledge discussions with Dipankar
Bhattacharya, Pawel Haensel and Mitja Rosina. The author thanks the Centre for
Theoretical Physics, Jamia Millia Islamia and ICTP, Trieste for hospitality.

\section*{Data Availability}

No data was generated in this work.



\bibliography{refs} 




\label{lastpage}
\end{document}